\documentclass[%
 reprint,
superscriptaddress,
 amsmath,amssymb,
 aps,
]{revtex4-2}
\usepackage{graphicx}
\usepackage{dcolumn}
\usepackage{bm}

\usepackage{xcolor}
\usepackage{datatool}
\usepackage{mathtools}
\usepackage{amssymb}   
\usepackage{amsmath}
\usepackage{verbatim}   
\usepackage{upgreek}
\usepackage[utf8]{inputenc}
\usepackage{subcaption}

\newcommand\xep{\ensuremath{\mathrm{Xe^+}}}

\renewcommand\d{\ensuremath{\mathrm{d}}}
\newcommand\p{\ensuremath{\mathrm{p}}}

\newcommand{\intensity}[2]{$#1\times10^{#2}\mbox{ Wcm}^{-2}$}
\newcommand{\fs}[1]{#1~fs}
\newcommand{\as}[1]{#1~as}

\newcommand{\nm}[1]{#1~nm}
\newcommand{\au}[1]{#1~a.u.}
\newcommand{\evolt}[1]{#1~eV}
\newcommand{\etal}{\textit{et al.\ }}
\newcommand{\xuv}{\text{XUV}}
\newcommand{\wXUV}{$\omega_\xuv$}
\newcommand{\dipole}{$\mathbf{d}($\wXUV$, \tau)$}
\newcommand{\wIR}{$\omega_{\text{NIR}}$}

\begin{document}
\preprint{APS/123-QED}

\title{Phase evolution of strong-field ionization}

\author{Lynda Hutcheson}%
 \email{lhutcheson02@qub.ac.uk}
\affiliation{Centre for Light Matter Interaction, School of Mathematics and Physics, Queen's University Belfast, Belfast  BT7 1NN, United Kingdom}%

\author{Maximilian Hartmann}
\affiliation{Max-Planck-Institut für Kernphysik, Saupfercheckweg 1, 69117 Heidelberg, Germany}

\author{Gergana D. Borisova}
\affiliation{Max-Planck-Institut für Kernphysik, Saupfercheckweg 1, 69117 Heidelberg, Germany}

\author{Paul Birk}
\affiliation{Max-Planck-Institut für Kernphysik, Saupfercheckweg 1, 69117 Heidelberg, Germany}

\author{Shuyuan Hu}
\affiliation{Max-Planck-Institut für Kernphysik, Saupfercheckweg 1, 69117 Heidelberg, Germany}

\author{Christian Ott}
\affiliation{Max-Planck-Institut für Kernphysik, Saupfercheckweg 1, 69117 Heidelberg, Germany}

\author{Thomas Pfeifer}
\affiliation{Max-Planck-Institut für Kernphysik, Saupfercheckweg 1, 69117 Heidelberg, Germany}

\author{Hugo W. van der Hart}
\affiliation{Centre for Light Matter Interaction, School of Mathematics and Physics, Queen's University Belfast, Belfast  BT7 1NN, United Kingdom}

\author{Andrew C. Brown}
\email{andrew.brown@qub.ac.uk}
\affiliation{Centre for Light Matter Interaction, School of Mathematics and Physics, Queen's University Belfast, Belfast  BT7 1NN, United Kingdom}%


\date{\today}

\begin{abstract}
We investigate the time-dependent evolution of the dipole phase shift induced by strong-field ionization (SFI) using attosecond transient absorption spectroscopy (ATAS) for time-delays where the pump-probe pulses overlap. We study measured and calculated time-dependent ATA spectra of the ionic $4\d\rightarrow5\p$ transition in xenon, and present the time-dependent line shape parameters in the complex plane. We attribute the complex, attosecond-scale dynamics to the contribution of three distinct processes: accumulation of ionization, transient population, and reversible population of excited states arising from polarization of the ground state.
\end{abstract}
\date{\today}

\maketitle
At the heart of attosecond science lies strong field ionization
(SFI)~\cite{SFI_review2005,SFA_review}. Not only is SFI a fundamental process vital for our
understanding of the purely quantum effect of tunneling through the
barrier~\cite{Yudin2001a}, it is also responsible for phenomena
such as above-threshold ionization~\cite{Agostini1979} and high harmonic generation~\cite{Corkum1993,lewenstein}, the latter of which has been harnessed to generate coherent
attosecond pulses~\cite{Kienberger2004,Gaumnitz2017}. 

Early investigations of SFI dynamics using attosecond light sources measured the ionization yield~\cite{Uiberacker2007}. For instance, Ne atoms were irradiated by a superposition of a strong near-infra-red (NIR) light field, and an ultra-short attosecond pulse. Measurement of the Ne$^{2+}$ yield as a function of delay time between the two pulses gives insight into the evolution of the ionization yield generated by the NIR pulse.

Since these early studies of ionization yield, investigating the phase behavior of ionized systems has become commonplace through, e.g. measurement of photoionization time delays~\cite{delay_photoemission,Kluender2011,Schultze2011,joseph2020angle,patchkovskii2023theory}.  However, these studies focus on the phase of the photoelectrons, and not on the phase evolution during ionization. 

Attosecond transient absorption spectroscopy (ATAS) has been used to investigate SFI in `real time'. 
By leveraging the charge-state specificity~\cite{Kobayashi2018} and site-selectivity~\cite{Kobayashi2019} of
the transitions, alongside the use of ultrashort probe pulses, ATAS enables the study of ionization dynamics with remarkable precision on the attosecond time scale~\cite{Beck2015,golubev2021core,Chen2012}.
Recent advances in spectrometer resolution have enabled investigations of ATAS absorption lines with respect to both their amplitude and their specific shape~\cite{Ott2013}. This provides access to the dipole phase that, in the context of SFI, can be related to the time evolution of the residual ion. Retrieving the time-dependent phase allows the reconstruction of time-dependent electron wavepackets~\cite{Wirth2011,Goulielmakis2010}.  

At high NIR intensities (in the over-the-barrier regime) the interactions between the residual ion and the outer electron were found to be negligible~\cite{Pabst2012, Wirth2011}.
The observed dipole phase behavior was instead attributed to the `field-induced
mixing of the excited $N$-electron states with the neutral ground state'. However, for lower-intensity, multi-cycle NIR pulses, we expect the coupling between the residual ion and the outgoing electron to increase in importance.

\begin{figure}
    \begin{subfigure}[t]{0.49\linewidth}
        \centering
        \includegraphics[height=1in]{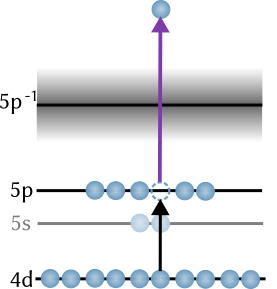}
        \caption{Continuum}\label{fig:mechanism_ionised} \end{subfigure}%
    \begin{subfigure}[t]{0.49\linewidth}
        \centering
        \includegraphics[height=1in]{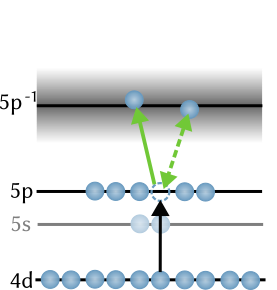}
        \caption{Bound}\label{fig:mechanism_bound}
    \end{subfigure}
    \captionsetup{justification=raggedright}
    \caption{Schematic diagram of the proposed mechanism for strong-field ionization in the Xenon atom: an NIR-pump pulse drives SFI of
    a $5\p$ electron either (a) directly into the continuum---purple arrow---or
    (b) into bound states lying close to the ionization threshold---solid green arrow. The excited bound states can also be populated reversibly via polarization of the ground state (b, dashed green arrow). The
    characteristics of the $5\p$ hole
 are captured through the promotion of a $4\d$
    electron into the vacancy by a time-delayed XUV-probe pulse (black
    arrow).}\label{fig:mechanism}
\end{figure}

In this paper, we use ATAS to investigate the time evolution of the dipole phase during SFI of Xe. Figure~\ref{fig:mechanism_ionised} sketches the
SFI mechanism under consideration. An intense,
\nm{760} NIR pulse (FWHM \fs{4.5}) drives SFI of a valence
$5\p$ electron. A time-delayed extreme ultraviolet (XUV) pulse (FWHM \as{250})
then probes the target by exciting a core $4\d$ electron into the now-vacant $5\p$ hole. At lower intensities, this hole will be affected by the outgoing electron.
To investigate this, we
analyze ATA spectra from experiment and 
R-Matrix with Time-dependence (RMT) calculations~\cite{RMT_cpc}. Details of the RMT calculations/experimental set-up can be found in
Refs.~\cite{hartmann2022core,Stooss2019}.

\begin{figure}
\includegraphics[width=0.98\linewidth]{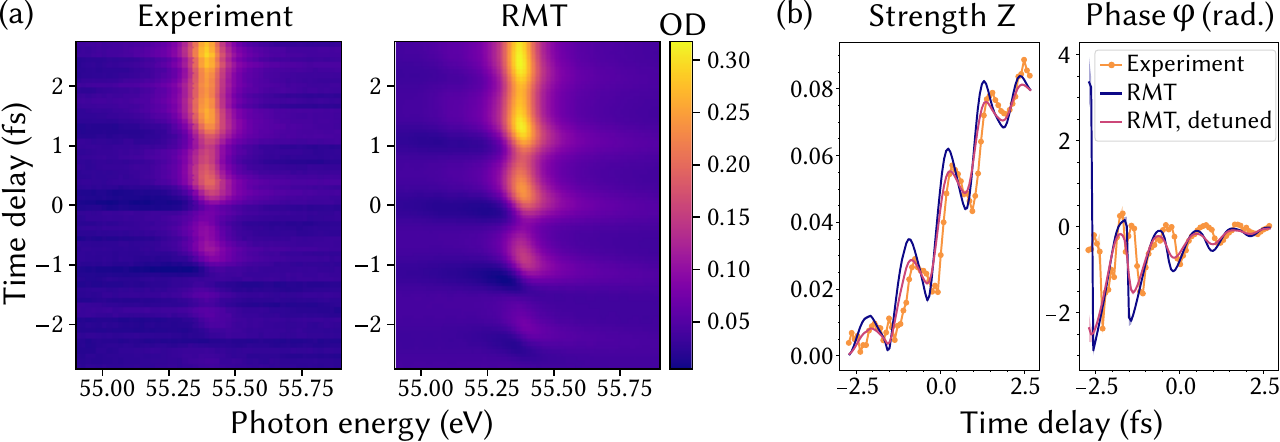}
\captionsetup{justification=raggedright}
\caption{(a) ATAS time-delay scans (delay of
XUV-probe with respect to NIR-pump, centred on \fs{0}) for the $4\d\rightarrow5\p$ transition in Xenon for both experiment (left, NIR intensity \intensity{0.9}{14}) and \textit{ab initio} RMT calculations (right, NIR intensity \intensity{0.8}{14}).
The NIR intensities are chosen such that the ionization yield matches. (b) Fit results
of the lineshape model~\cite{Pabst2012} to the absorption spectrum to obtain both the strength $z$ (left) and dipole phase $\varphi$ (right) for experiment and two separate RMT calculations: one with the central XUV energy tuned to the $4\d\rightarrow5\p$ resonance, and one with energy detuned by
\evolt{3}.}\label{fig:OD}
\end{figure}

RMT solves the time-dependent Schrödinger equation (TDSE)
for multi-electron systems driven by strong, arbitrarily polarised
laser pulses. This yields the time-dependent wavefunction, from which we may compute the time-dependent expectation value of the dipole operator, $\mathbf{d}(t, \tau)$, for pump-probe time delay $\tau$. The absorption spectrum can then be calculated as:
\begin{equation}
    \label{eq:ATAS_equation}
    \sigma(\omega, \tau) = 4\pi\alpha\omega \text{Im}\left[ \frac{\tilde{\mathbf{d}}(\omega, \tau)}{\tilde{\mathbf{E}}(\omega, \tau)} \right],
\end{equation}
where $\omega$ is the photon energy, $\alpha$ is the fine-structure constant and $\tilde{\mathbf{d}}(\omega, \tau)$ and $\tilde{\mathbf{E}}(\omega, \tau)$ are the Fourier transforms of the time-dependent dipole moment, $\mathbf{d}(t, \tau)$, and the applied electric field, $\mathbf{E}(t, \tau)$, respectively. We calculate (or measure) the absorption spectrum around the $4\d\rightarrow5\p$ transition energy, \evolt{55.39}, as shown in Figure.~\ref{fig:OD}a.

The delay-dependent amplitude, $z$, and phase $\varphi$ of the dipole response are extracted from the ATA spectra through a fitting procedure~\cite{Pabst2012} for each time delay $\tau$. The results of this fitting procedure are shown in Figure.~\ref{fig:OD}b. The model has been used previously in the analysis of this ionic core-to-valence
transition in Xe to examine the absorption strength~\cite{Sabbar2017} and phase~\cite{hartmann2022core} independently. In this work, we examine them simultaneously.

Kaldun \textit{et al.}~\cite{Kaldun2014} presented a scheme to quantify and observe the amplitude and phase changes of excited states coupled by a laser field. Relative phase and amplitude modifications of the excited states are revealed in the complex-valued dipole response.
Therefore, instead of considering the real-valued amplitude, $z$, and phase, $\varphi$, of
the dipole response independently---as in Fig.~\ref{fig:OD}b---we can investigate them together in the complex plane. The dipole takes the form
\begin{equation}
    \label{eq:complex_fit}
   \mathbf{d}(\omega_\xuv, \tau) \propto z(\tau) e^{i\varphi(\tau)}.
\end{equation}
Here, \dipole\ denotes the frequency-dependent dipole evaluated at the energy of
the XUV pulse, \wXUV\ for time-delay $\tau$. For ease of notation, we
ignore proportionality constants and consider the quantities equal. Kaldun
\etal~\cite{Kaldun2014} looked at autoionizing states, and transitions involving valence electrons. Here, we are looking at core transitions, only accessible following ionization or whilst a valence electron undergoes emission.

\begin{figure}[t]
    \centering
    \includegraphics[width=0.99\linewidth]{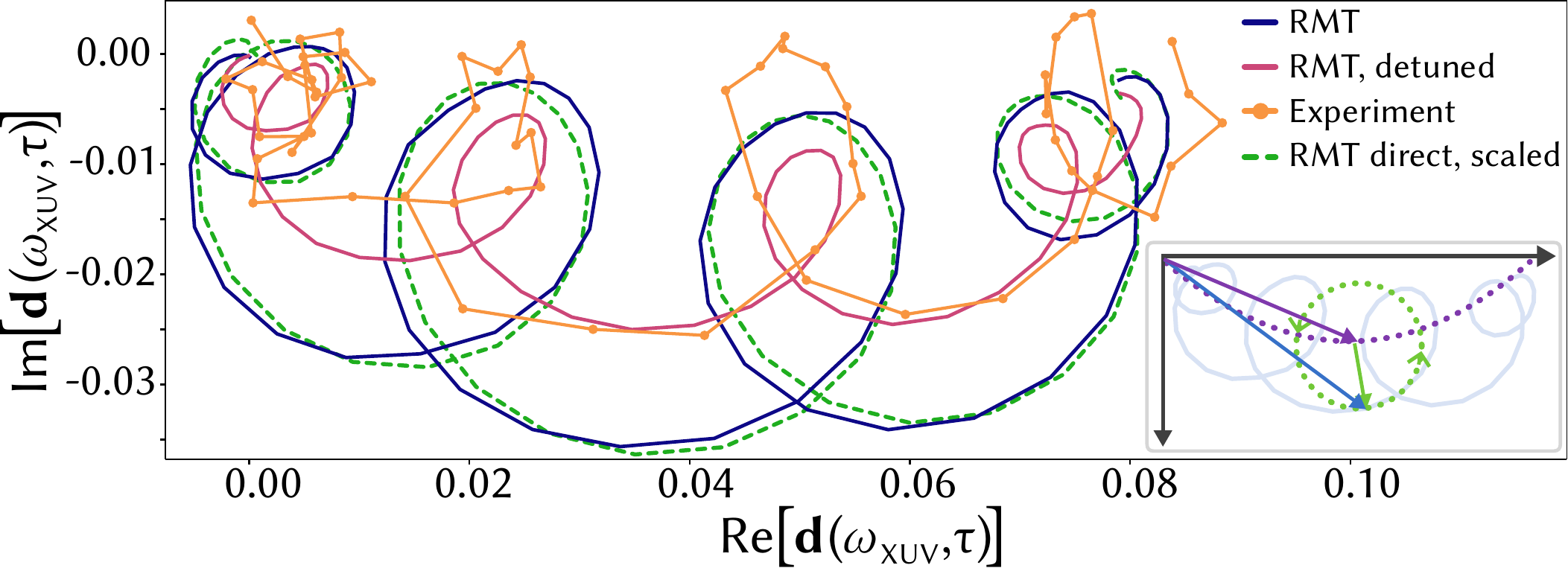}
    \captionsetup{justification=raggedright}
    \caption{The complex dipole response, \dipole, extracted from the absorption
    line shapes shown in figure \ref{fig:OD},
 including the complex dipole response extracted directly from RMT calculations. (Inset) A sketch of the two components comprising the overall behavior
    (blue): movement of a center-point (the `dip', purple) with a superimposed rotation
    (the `loops', green).}\label{fig:complex}
\end{figure}

The dipole response, \dipole, obtained from both RMT and experimental absorption
spectra are plotted in Fig.~\ref{fig:complex}. 
The behavior has two components, as shown on the inset of Fig.~\ref{fig:complex}:
an overall `dip' into the lower half-complex plane with superimposed rotation (`loops'). We obtain excellent qualitative and quantitative agreement between
experiment and theory: the response is of the same
magnitude, and the position of the loops is similar. 

To explain this response, we propose that SFI does not drive the $5\p$ electron into a well-defined continuum state, as might be
expected with photoionization, for instance. Rather, as depicted in
Fig.~\ref{fig:mechanism}, in addition to ionization (Fig.~\ref{fig:mechanism_ionised}) the IR populates several intermediate states (Fig.~\ref{fig:mechanism_bound}). 
These comprise excited states, representing the time-dependent polarization of Xe, 
(quasi-)bound states such as high-lying Rydberg states~\cite{Nubbemeyer2008}, or transient low-energy continuum states. The probed, $5\p$ hole thus exists in a superposition of states.

\begin{figure}
    \centering
    \includegraphics[width=0.95\linewidth]{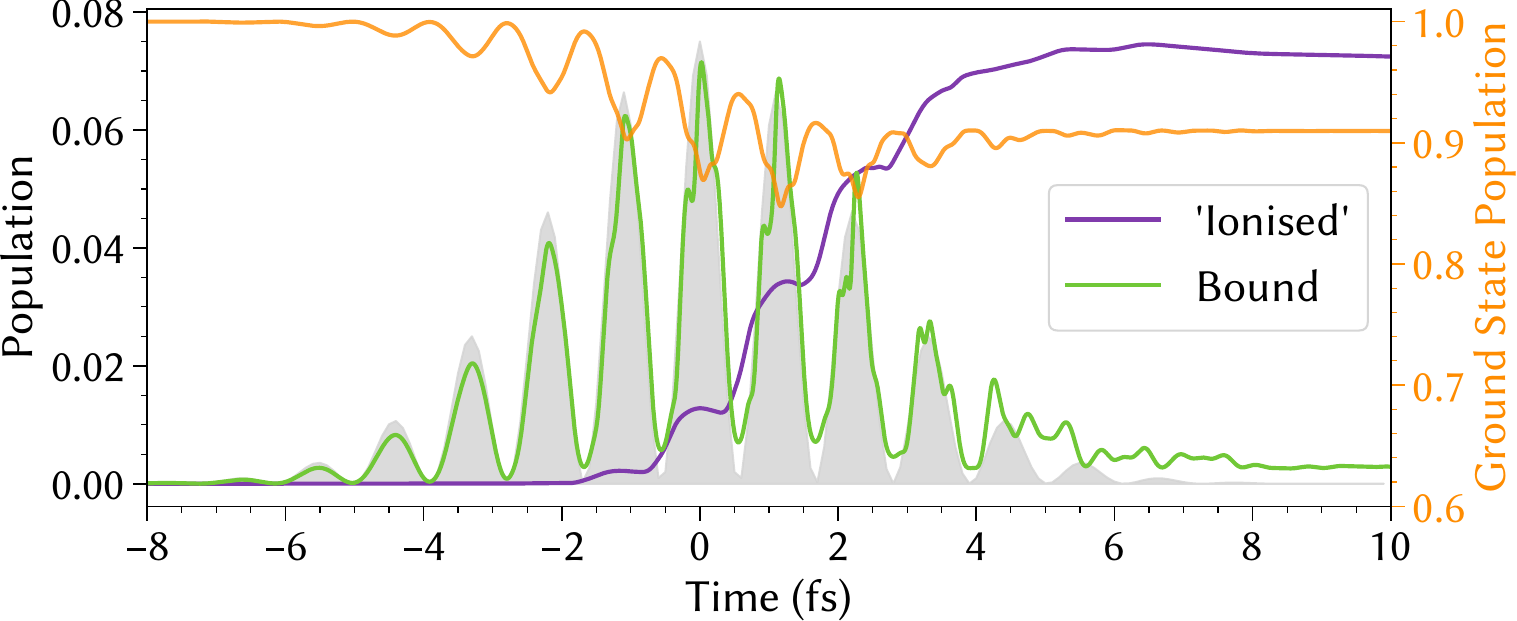}
    \captionsetup{justification=raggedright}
    \caption{
    Populations of various states during SFI ionization as extracted from an RMT calculation
    using only the NIR pulse, centered at \fs{0}, for \intensity{0.8}{14} intensity (gray background, not to scale). 
    Shown is the population of the ground state (solid orange line), of a set of excited $^1 P^o$ bound
    states (green line) and the total population beyond a radial distance of 
    \au{20} (purple line) which is used to give an indication of the ionization yield. The rise in population beyond \au{20} is time-delayed with respect to the peak of the NIR pulse
as the outgoing wavepacket takes time to propagate outwards.
    }\label{fig:populations}
\end{figure}

To support this proposed mechanism, we extract the
populations of the bound states as the atom undergoes SFI 
(Fig.~\ref{fig:populations}) from an RMT calculation which includes only the NIR pulse. 
For simplicity, the bound-state population in Fig.~\ref{fig:populations} is restricted to a set of near-threshold bound states in the $^1 P^o$ symmetry.
The ground-state behavior consists of two components: an overall decrease-- corresponding to ionization and excitation-- and oscillations at twice the driving NIR frequency, \wIR--
indicative of dynamic polarization~\cite{Chini_stark_shift,Sabbar2017} or the coupling of two near-resonant quantum states~\cite{Kaldun2014,Blatterman_ATAS}. The excited bound-state population also oscillates with 2\wIR, but, importantly, does not return to zero at the minima of the oscillations. This is indicative of a transient excited-state population beyond what can be explained through ground-state polarization.

Further evidence can be gathered from the wavefunction density.
Towards the peak of the pulse, the population beyond \au{20} 
begins to rise in a stepwise
fashion. This population is not the instantaneous
ionization yield: it comprises both ionized and bound electrons, as Xe supports
several, high-lying Rydberg states that extend beyond \au{20}. Overall, we thus confirm within the RMT calculation that (highly) excited states of the neutral atom are populated, in agreement with previous findings~\cite{Nubbemeyer2008}.

\begin{figure*}[htb]
    \centering
    \includegraphics[width=0.99\linewidth]{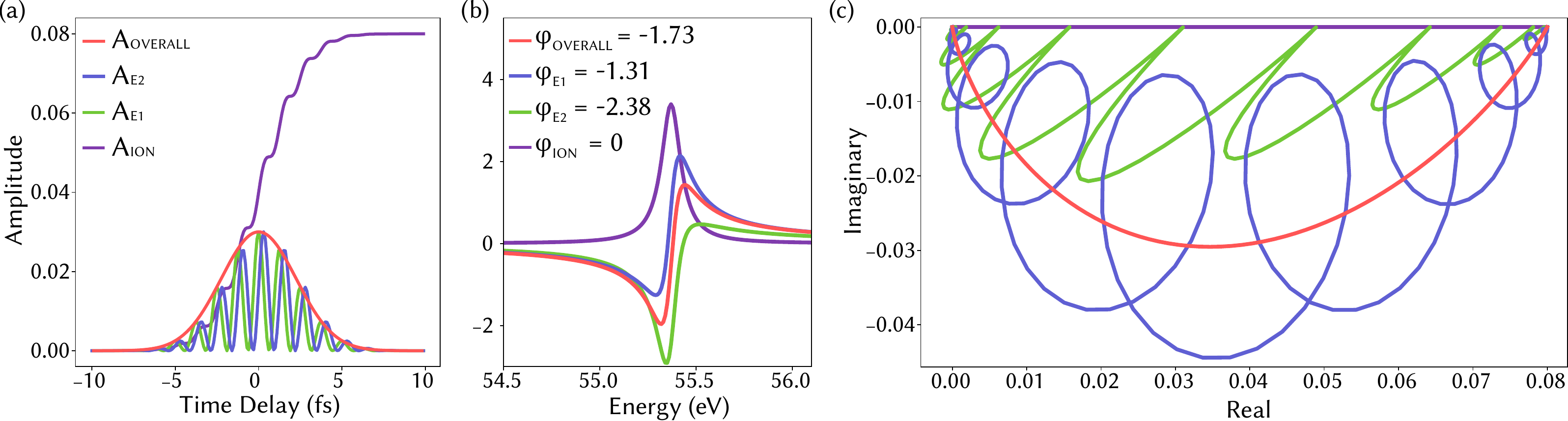}
    \captionsetup{justification=raggedright}
    \caption{The details of the simple model for the dipole response. We
        model three channels: an effective ionization channel (purple) and
        two channels representing excited transient Rydberg states (green and blue). In reality, there are many more. We also consider an `overall' excitation channel (red line). (a) Time-dependent amplitudes considered in the model. For the ionized
        channel, the integral of the squared NIR pulse is used.
        For the bound channels, the squared NIR pulse is used, with a time delay between the two channels. For the `overall' excitation channel, we use the NIR pulse envelope. Each effective channel is assigned a constant dipole phase. The corresponding absorption line shape for each phase is shown in (b). 
        (c) The resultant complex dipole response of the coherent sum between various channels showing: the ionization channel only (purple), the ionization channel plus one effective excitation channel (green), the ionization channel plus both excitation channels (blue), and the `overall excitation' neglecting the 2\wIR\ oscillations (red line).}\label{fig:model}
\end{figure*}

To demonstrate the link between the strong-field excitation and ionization, we consider a minimal model comprising the coherent dipole response of just three channels: one effective ionization channel and two channels to model transient excited (Rydberg) states. The coherent dipole response is defined by:
\begin{equation}
    \mathbf{d}(\tau) = z_{\text{ION}}(\tau)e^{i\varphi_{\text{ION}}} + z_{\text{E1}}(\tau)e^{i\varphi_{\text{E1}}} + z_{\text{E2}}(\tau)e^{i\varphi_{\text{E2}}}
\end{equation}
where $z$ and $\varphi$ are the amplitude and phase of each effective channel.

The time-dependent amplitude of each channel (Fig.~\ref{fig:model}a) is derived from the NIR
pulse profile. We model the ionization channel amplitude, $z_{\text{ION}}$, as the cumulative integral of the squared NIR pulse and the population in excited states ($z_{\text{E1}}$ and $z_{\text{E2}}$) is modeled with
the squared NIR pulse. These time-dependent amplitudes align with
populations obtained from RMT calculations involving just the NIR pulse (Fig.~\ref{fig:populations}). 

To each of these channels, we assign a dipole phase. While the dipole phase of each channel remains unchanged, when we coherently sum the channels with the different time-dependent amplitudes, the phase of the coherent mixture changes. The SFI channel has a phase of $\varphi_{\text{ION}}=0$ representing an outgoing electron uncoupled with the $5\p$ hole. Considering only this channel, the observed dipole response remains strictly real, growing in magnitude with the ionization yield (Fig.~\ref{fig:model}c).
Adding a second channel for strong-field excitation with a non-zero dipole phase produces oscillations in the dipole response, but fails to reproduce the observed behavior, particularly the characteristic loops and dip.

To capture the full behavior, we introduce a third excitation channel. The two effective excitation channels must have different, non-zero, dipole phases (in Fig.~\ref{fig:model} we use $\varphi_1 = -2.38$, $\varphi_2=-1.31$) and be time-delayed with respect to each other. Obtaining loops of the correct orientation and magnitude requires this time delay to be \as{$300\pm50$}. When we neglect dynamic polarization by ignoring the 2\wIR\ oscillations and modeling the time-dependent amplitude of the channels using the NIR envelope instead,
the dipole response exhibits the correct dip structure, however without loops.

\begin{figure}[htb]
\includegraphics[width=0.99\linewidth]{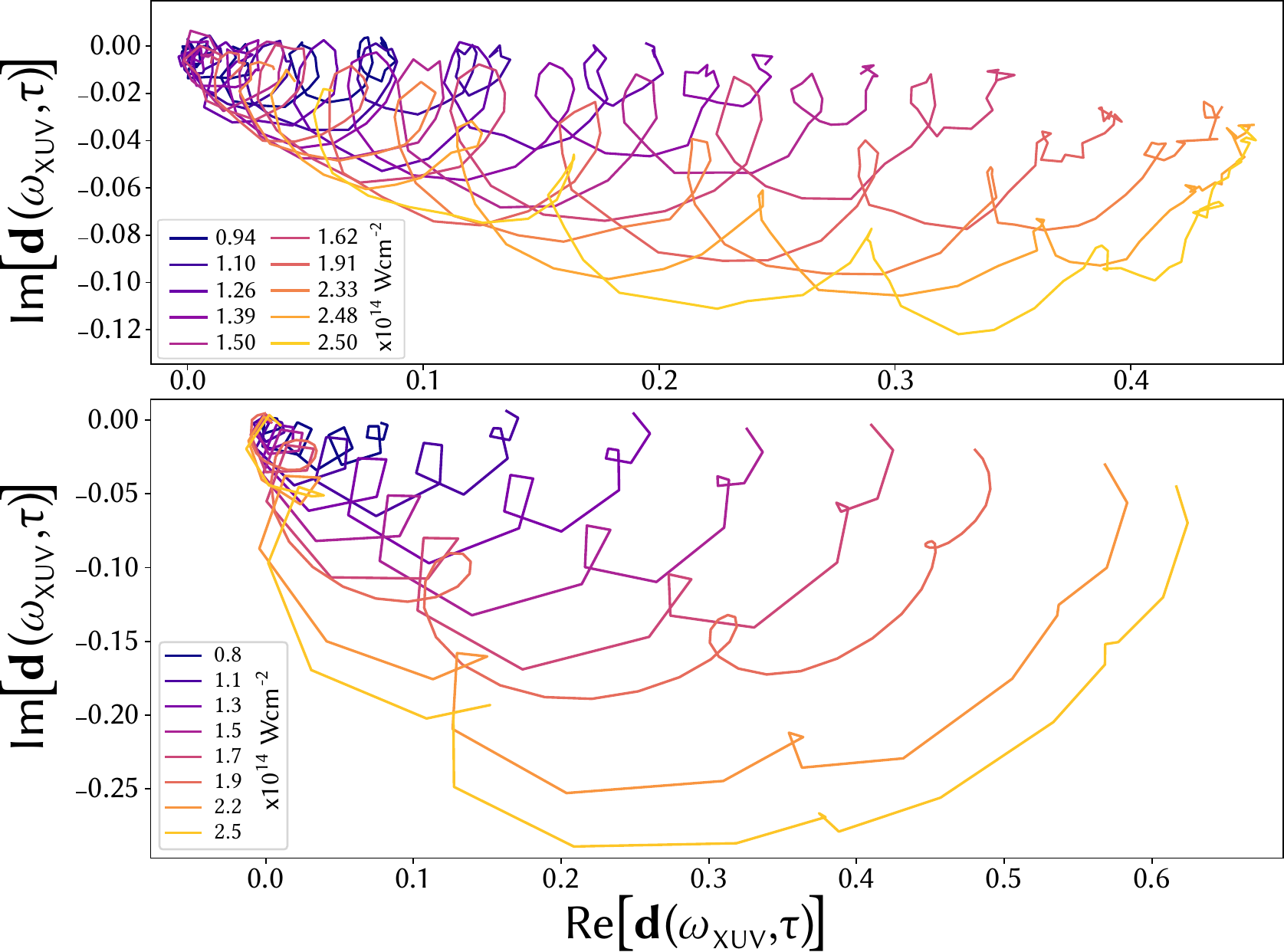}
\captionsetup{justification=raggedright}
\caption{The complex dipole response, \dipole, extracted from absorption line shapes obtained from both experiment (top) and RMT (bottom) for increasing NIR intensities as denoted by the legends.}\label{fig:complex_intensity}
\end{figure}

Interestingly, this is the behavior we observe at high NIR intensity, \intensity{2.5}{14}.
Figure~\ref{fig:complex_intensity} shows that as NIR intensity increases, the overall magnitude of the dipole response grows while the loop size decreases.
At higher NIR intensities, the direct SFI pathway will
dominate over any excitation into transient states, limiting the visibility of the interference effects. Or, adopting a simple view of the dynamics, the
continuum electron will attain a higher energy due to the stronger field, thus
restricting its opportunity to interfere with the bound electrons.
This dominance of SFI at higher intensities explains the intensity-dependence of the
oscillations in the dipole phase reported in
Ref.~\cite{hartmann2022core} and justifies neglecting the interactions between the residual ion and the outer electron at high NIR intensities, as in Ref.~\cite{Pabst2012}. All of this implies that the rich dynamics, including interference between multiple excitation channels, are supported only at moderate intensities, \intensity{<2}{14}.

This interpretation also explains the effect of detuning the broadband XUV energy from the resonant
$4\d\rightarrow5\p$ transition (Figure~\ref{fig:complex}). The real part of the dipole response is largely unaffected, being dominated by the \xep ion response. However, the imaginary part decreases notably in magnitude, most clearly seen in a reduction of the size of the loops. 
The model calculation implies that the size of the loops is mediated by the relative contribution of multiple excited states.
By detuning the XUV, we modify the relative contribution of the several $5\p$ hole states to the absorption. 
The reduced loop size suggests decreased interference between the different excitation pathways, which can be expressed through a time delay, as in the minimal model.

All of this confirms our proposed mechanism; the NIR drives a
combination of SFI and strong field excitation. The XUV then probes the $5p$ hole created by these processes, and the dipole response encodes the interference between the excited and ionized states. The dynamic polarization yields 
2\wIR\ oscillations in the excited state populations, leading to loop structures in the complex dipole response (Fig.~\ref{fig:complex}). The overall
structure, including the `dip' into the lower half of the complex plane, can be
explained by considering the time-dependent coherent mixing between the SFI and
strong-field excitation pathways. Importantly, the full dynamics are mediated by interference between multiple excitation pathways. At high NIR intensity (\intensity{>2}{14}) these interference effects are washed out by the direct ionization pathway.

In conclusion, we have carried out an analysis of ATA spectra derived from RMT calculations including a comparison with experimental results. Presentation of time-dependent ATAS line shape parameters in the complex plane helps illustrate and quantify the interfering dynamics encoded within the absorption throughout the NIR pulse. Few-cycle and low-intensity NIR pulses lead to complex dynamics during strong-field ionization, quantified through the attosecond XUV absorption, where we have identified the contribution of three distinct processes: accumulation of ionization, transient population of excited states, and reversible population of excited states arising from polarization of the ground state.

Quite generally, strong-field ionization at intermediate intensities is the most demanding region for a precise physical understanding, as the interplay between the atomic structure and the strong field needs to be described accurately. The excellent agreement between theory and experiment provides further evidence that we have a full range of techniques to accurately explore physics in this demanding regime. In the future, extending the scope of phase-resolved strong-field ionization holds promise to the measurement and control of real-time attosecond entanglement dynamics of the ionizing electron and the residual ion in both atoms and molecules~\cite{Knoll_entanglement,vrakking_entanglement,laurell2023measuringquantumstatephotoelectrons,shobeiry2024emission}, with precision phase- and quantum-state-resolved attosecond information.

This work benefited from computational support by CoSeC, the Computational Science Centre for Research Communities, through CCPQ. ACB and HWvdH acknowledge funding from the EPSRC under Grants No. EP/T019530/1, EP/V05208X/1, and EP/R029342/1. This work relied on the ARCHER UK National Supercomputing Service, for which access was obtained via the UK-AMOR consortium funded by EPSRC. We also acknowledge support by the Deutsche Forschungsgemeinschaft (DFG, German Research Foundation) under Germany’s Excellence Strategy EXC2181/ 1-390900948 (the Heidelberg STRUCTURES Excellence
Cluster) and by the European Research Council (ERC) (X-MuSiC 616783).

\bibliography{mybib.bib}
\end{document}